\documentclass[pra,11pt,onecolumn, showkeys,showpacs]{revtex4}

\usepackage{times}

\begin{document}

\title{Comment on "proactive quantum secret sharing"}
\author{Gan Gao\footnote{Corresponding author. E-mail:
gaogan0556@163.com}}

\affiliation{Department of Electrical Engineering, Tongling
University, Tongling 244000, China}

\date{\today}

\begin{abstract}
In the paper [Quantum Inf. Process. {\bf 14,} 4237-4244 (2015)], Qin and Dai proposed a proactive quantum secret sharing scheme. We study the security of the proposed scheme and find that it is not secure. In the distribution phase of the proposed scheme, two dishonest participants may collaborate to eavesdrop the secret of the dealer without introducing any error. Incidentally, a possible improvement of the distribution phase is given.
\end{abstract}



\maketitle

\noindent {\bf 1. Introduction }\\

As we all know, quantum secret sharing (QSS) is a counterpart of classical secret
sharing in quantum world. In QSS, a secret will be split into several pieces by
a dealer, and each participant holds a piece, and no subset of participants is
sufficient to extract the secret, but the entire set is. Note that not
all of the participants in QSS are credible, moreover, a
dishonest participant has more power to attack a QSS scheme
than a outside eavesdropper. Therefore, we should pay more attention to the attack of
dishonest participant(s) when analyzing the security. Recently, by using Bell states and unitary operations, Qin and Dai [1] proposed
a proactive QSS scheme, which has the three phases: the distribution phase, the updating phase and the recovery phase. The Qin-Dai QSS scheme [1] is interesting
because all participants in it can periodically update their key shares, meanwhile,
the secret of the dealer is changeless. However, it is somewhat a pity that the Qin-Dai QSS scheme
has a drawback of security. That is, in the distribution phase of the scheme, the two dishonest participants (the first participant and the last participant) may collaborate to eavesdrop the secret of the dealer without being detected. \\

\noindent {\bf 2. Distribution phase in the Qin-Dai QSS scheme}\\

Before describing our attack strategy, let us review the distribution phase in the Qin-Dai QSS scheme as follows:
(1) Alice generates a Einstein-Podolsky-Rose (EPR) pair sequence, which is denoted with $\{|\Psi\rangle_{x_{1},y_{1}},|\Psi\rangle_{x_{2},y_{2}},...,|\Psi\rangle_{x_{m},y_{m}}\}$ (here, $x_{i},y_{i}\in \{0,1\}$). Each EPR pair is randomly in one of the four Bell states: $|\Psi_{00}\rangle=(|00\rangle+|11\rangle)/\sqrt{2}, |\Psi_{01}\rangle=(|00\rangle-|11\rangle)/\sqrt{2}, |\Psi_{10}\rangle=(|01\rangle+|10\rangle)/\sqrt{2}, |\Psi_{11}\rangle=(|01\rangle-|10\rangle)/\sqrt{2}$. (2) Alice takes the first particle and the second particle from each EPR pair to form [x] sequence and [y] sequence, respectively. (3) Alice prepares some decoy particles for eavesdropping detection. (4) Alice inserts the decoy particles into the [y] sequence. Then she sends the [y] sequence including decoy particles to Bob$_{1}$. (5) After confirming that Bob$_{1}$ receives the sequence, Alice and Bob$_{1}$ start to check the security of the [y] sequence transmission by the decoy particles. (6) Bob$_{1}$ encodes his private key by performing the unitary operation $U_{u^{1}_{i},v^{1}_{i}}$ on the $i$th particle in the [y] sequence. (7) Bob$_{1}$ inserts some decoy particles, which are prepared in advance, into the [y] sequence. Next, he sends the [y] sequence to Bob$_{2}$. Similarly, Bob$_{2}$ and Bob$_{1}$ check the security of the [y] sequence transmission by the decoy particles. Then, Bob$_{2}$
does the similar action as Bob$_{1}$, and this process is continued until Bob$_{n}$. (8) Bob$_{n}$ sends the [y] sequence to Alice. Here, the security of the sequence transmission is checked by the decoy particles that Bob$_{n}$ inserts into the [y] sequence. (9) The [y] sequence and the [x] sequence are gotten back to the EPR pair sequence by Alice. Then she performs a Bell state measurement on each EPR pair, and her measurement outcomes are denoted with $\{|\Psi\rangle_{x'_{1},y'_{1}},|\Psi\rangle_{x'_{2},y'_{2}},...,|\Psi\rangle_{x'_{m},y'_{m}}\}$. (10) Alice extracts the secret: ($x_{1}\oplus x'_{1},y_{1}\oplus y'_{1},x_{2}\oplus x'_{2},y_{2}\oplus y'_{2},...,x_{m}\oplus x'_{m},y_{m}\oplus y'_{m}$).\\

\noindent {\bf 3. Insecurity of the Qin-Dai QSS scheme}\\

In this section, we start to describe our attack strategy in detail. By the way, this attack strategy will be implemented by the first participant (Bob$_{1}$) and the last participant (Bob$_{n}$). In other words, Bob$_{1}$ and Bob$_{n}$ can obtain Alice's secret without help of the other participants
by this attack strategy. This attack strategy is described as follows:

In advance, Bob$_{1}$ prepares a sequence composed of $m$ EPR pairs, in which each pair is in $|\psi\rangle_{ab}=(|01\rangle-|10\rangle)/\sqrt{2}$, and takes out $a$ and $b$ particles from each pair to form [a] sequence and [b] sequence, which is also known by Bob$_{n}$ because they are a cahoot [2,3]. After the [y] sequence is  received by Bob$_{1}$, firstly the security of the sequence transmission is checked, and then he performs the $U_{u^{1}_{i},v^{1}_{i}}$ operation on the $i$th particle in the [y] sequence, and secretly sends the [y] sequence to Bob$_{n}$. Next, Bob$_{1}$ inserts some decoy particles into the [b] sequence, and sends the [b] sequence, instead of the [y] sequence, to Bob$_{2}$. After Bob$_{2}$ receives the sequence, he also checks the security of the [b] sequence transmission with the decoy particles. Since Bob$_{2}$ can't detect that Bob$_{1}$ has played a replacing trick, he performs his secret unitary operations as of old. Obviously, now Bob$_{2}$'s unitary operations are not performed on particles in the [y] sequence, but on particles in the [b] sequence, which is not known by Bob$_{2}$. After encoding, Bob$_{2}$ inserts some decoy particles into the [b] sequence and sends the sequence to Bob$_{3}$. According to Step (7), Bob$_{3}$,...,Bob$_{n-1}$ will do the similar action as Bob$_{2}$. Then, Bob$_{n-1}$ sends the [b] sequence to Bob$_{n}$. In order not to be detected, Bob$_{n}$ also checks the security of the [b] sequence transmission with Bob$_{n-1}$'s inserting decoy particles. Now, Bob$_{n}$ holds the [b] sequence and Bob$_{1}$ holds the [a] sequence. Next, the two dishonest participants perform Bell state measurement on each EPR pair. Since each EPR pair that Bob$_{1}$ previously prepared is in $|\psi\rangle_{ab}$, by comparing the Bell state measurement outcome with $|\psi\rangle_{ab}$, Bob$_{1}$ and Bob$_{n}$ easily obtain $U_{i}=U_{u^{n-1}_{i},v^{n-1}_{i}}U_{u^{n-2}_{i},v^{n-2}_{i}}\cdot\cdot\cdot U_{u^{2}_{i},v^{2}_{i}}$. Afterwards, Bob$_{n}$ immediately performs the $U_{u^{n}_{i},v^{n}_{i}}U_{i}$ operation on the $i$th particle in the [y] sequence, and inserts some decoy particles into the [y] sequence, and sends the [y] sequence including decoy particles to Alice. Since that Alice checks the security of the received sequence transmission only depends on Bob$_{n}$'s inserting decoy particles, she can not detect the eavesdropping of Bob$_{1}$ and Bob$_{n}$. From Step (10), we see that Alice's secret is shown with ($x_{i}\oplus x'_{i},y_{i}\oplus y'_{i}$). By simply calculating, ($x_{i}\oplus x'_{i},y_{i}\oplus y'_{i}$) is equal to ($u^{n}_{i}\oplus u^{n-1}_{i}\oplus\cdot\cdot\cdot\oplus u^{1}_{i},v^{n}_{i}\oplus v^{n-1}_{i}\oplus\cdot\cdot\cdot\oplus v^{1}_{i}$) in $U_{u^{n}_{i},v^{n}_{i}}U_{u^{n-1}_{i},v^{n-1}_{i}}\cdot\cdot\cdot U_{u^{1}_{i},v^{1}_{i}}$.

According to the above analysis, we see there is a security loophole in the distribution phase of the Qin-Dai QSS scheme, that is, Alice's secret can not be safely shared among all the participants. Next, we will give an improvement for the distribution phase of the Qin-Dai QSS scheme. In order to reserve the character of the distribution phase, we will only change the method that Alice checks the security of the [y] sequence transmission. In other words, our modifying content will focus on Step (8) in the distribution phase of the Qin-Dai QSS scheme. We modify Step (8) as follows: (8$'$) Bob$_{n}$ sends the [y] sequence to Alice. After confirming that Alice receives the [y] sequence, first, Bob$_{n}$ uses the decoy particles to check the security that the [y] sequence is transmitted from Bob$_{n}$ to Alice. Then, Alice randomly selects some EPR pairs, and randomly uses $\{|0\rangle, |1\rangle\}$ basis or $\{|+\rangle=(|0\rangle+|1\rangle)/\sqrt{2}, |-\rangle=(|0\rangle-|1\rangle)/\sqrt{2}\}$ basis to measure the $x$ particles in the selected EPR pairs. Next, she publishes the positions of the measured $x$ particles in the [x] sequence, and doesn't publish her measurement outcomes. Alice requires that each participant publishes the operations performed on the $y$ particles in the same positions. Note that, the orders that they publish the operations are random. After these, Alice uses a correct measuring basis to measure the corresponding $y$ particles. By comparing the deductive outcomes with the measuring outcomes, she can judge whether the whole quantum channel is secure.
\\

\noindent {\bf 4. Conclusion}\\

In conclusion, we have proposed a attack strategy on the Qin-Dai QSS scheme, and shown that, in the distribution phase of the scheme, Bob$_{1}$ and Bob$_{n}$ can collaborate to eavesdrop Alice's secret without introducing any error. In other words, we have proved that the Qin-Dai QSS scheme is not secure. We see that the distribution phase is the first phase in the Qin-Dai QSS scheme, and also most important phase. However, the interesting phase in the Qin-Dai QSS scheme should be the second phase, that is, the updating phase. For any quantum cryptography scheme, the security is essential. If the security can not be assured, even if the framework of the designed scheme is interesting, the designed scheme has not much value.
\\

\noindent {\bf Acknowledgements}\\

The author Gao thanks his parents for their encouragements. \\

\noindent {\bf References}

\noindent[1] Qin, H., and Dai, Y.: Proactive quantum secret sharing. Quantum Inf. Process. {\bf 14,} 4237-4244 (2015)

\noindent[2] Wang, T.Y., Wen, Q.Y., Gao, F., Lin, S., Zhu, F.C.: Cryptanalysis and improvement of multiparty quantum secret sharing schemes.
Phys. Lett. A {\bf 373,} 65-68 (2008)

\noindent[3] Gao, G.: Cryptanalysis of multiparty quantum secret sharing with collective eavesdropping-check.  Opt. Commun. {\bf 283} 2997-3000 (2010)

\enddocument